\documentclass[conference]{IEEEtran}

\IEEEoverridecommandlockouts
\usepackage{cite}
\usepackage{amsmath,amssymb,amsfonts}
\usepackage{algorithmic}
\usepackage{graphicx}
\usepackage{textcomp}
\usepackage{xcolor}
\def\BibTeX{{\rm B\kern-.05em{\sc i\kern-.025em b}\kern-.08em
    T\kern-.1667em\lower.7ex\hbox{E}\kern-.125emX}}
    
\usepackage{comment}

\usepackage{url}

\usepackage{todonotes}

\usepackage{siunitx}

\usepackage{float}

\usepackage{stfloats}

\usepackage{listings}

\usepackage{pgfplots, pgfplotstable,booktabs}
\usetikzlibrary{positioning}

\definecolor{codegreen}{rgb}{0,0.6,0}
\definecolor{codegray}{rgb}{0.5,0.5,0.5}
\definecolor{codepurple}{rgb}{0.58,0,0.82}
\definecolor{backcolour}{rgb}{0.95,0.95,0.92}

\lstdefinestyle{mystyle}{
    backgroundcolor=\color{backcolour},   
    commentstyle=\color{codegreen},
    keywordstyle=\color{magenta},
    numberstyle=\tiny\color{codegray},
    stringstyle=\color{codepurple},
    basicstyle=\small\normalfont\sffamily,
    breakatwhitespace=false,         
    breaklines=true,                 
    captionpos=b,                    
    keepspaces=true,                 
    numbers=left,                    
    numbersep=5pt,                  
    showspaces=false,                
    showstringspaces=false,
    showtabs=false,                  
    tabsize=2
}

\begin{document}

\bstctlcite{IEEEexample:BSTcontrol}

\title{PMT: Power Measurement Toolkit

}

 \author{

Stefano Corda$^{1}$, Bram Veenboer$^{2}$, Emma Tolley$^{1}$\\

\vspace{-0.4cm} \normalsize $^1$EPFL École Polytechnique Fédérale de Lausanne, $^2$ASTRON Netherlands Institute for Radio Astronomy \\\\

stefano.corda@epfl.ch, veenboer@astron.nl, emma.tolley@epfl.ch

}

\maketitle

\begin{abstract}

Efficient use of energy is essential for today's supercomputing systems, as energy cost is generally a major component of their operational cost. Research into ``green computing'' is needed to reduce the environmental impact of running these systems.
As such, several scientific communities are evaluating the trade-off between time-to-solution and energy-to-solution. While the runtime of an application is typically easy to measure, power consumption is not.

Therefore, we present the Power Measurement Toolkit (PMT), a high-level software library capable of collecting power consumption measurements on various hardware. The library provides a standard interface to easily measure the energy use of devices such as CPUs and GPUs in critical application sections.

\end{abstract}


\section{Introduction}
\label{sec:introduction}

Contemporary scientific applications such as particle hydrodynamics~\cite{10.1145/3394277.3401855}, radio-astronomical imaging~\cite{9709826} and Earth digital twin simulation~\cite{earth-digital-twin} have demanding compute requirements, in the order of the ExaFlops or more.
Moreover, top supercomputers~\cite{Dongarra2011} aim to reach Zettascale~\cite{liao_2018} performance in the current decade.
To achieve this goal, increasingly more powerful hardware is used. However, due to the end of Dennard's Scaling law, these advances in compute power come at the cost of a higher power consumption~\cite{6307773}.

Energy efficiency is a critical challenge in HPC (High-Performance Computing) as data centers are struggling to operate within stringent power budgets~\cite{10.1145/3388333.3403035} and also try to reduce their carbon footprint~\cite{9752341}.
Therefore, it is crucial to monitor the power consumption of HPC applications in order to evaluate trade-offs between achieved performance and energy efficiency~\cite{9188149}.

In this context, we present the Power Measurement Toolkit (PMT), a lightweight high-level library capable of collecting power measurements on various architectures, such as CPUs and GPUs. PMT can easily be used to probe specific application regions to measure power consumption and evaluate energy efficiency.


\section{Power Measurement Toolkit}
\label{sec:powermeasurement}

\begin{figure}[H] 
\centering
\includegraphics[width=8.5cm]{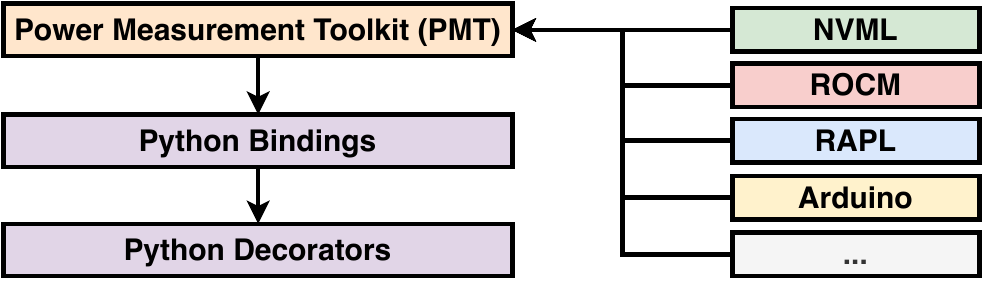}
\caption{Power Measurement Toolkit structure: the library interfaces with different power measurement back ends. The library also includes Python bindings and decorators to ease the measurement for application users.}
\label{fig:pmt}
\end{figure}

PMT's structure is reported in \emph{Fig.~\ref{fig:pmt}}. The library is written in C++ and is Linux-only; it interfaces with hardware vendors' API to collect power consumption. More precisely, for GPUs it uses NVML~\cite{nvml} for NVIDIA and rocm-smi~\cite{rocm} for AMD.
CPUs are monitored through the RAPL (Running Average Power Limit)~\cite{rapl} interface, or through LIKWID~\cite{5599200}.
PMT can profile other architectures, such as Xilinx FPGAs, and it can be easily extended to support new vendors' hardware.
Some other architectures expose their power usage information through files in sysfs (the \texttt{/sys} folder). PMT also has an interface to physical power sensors such as PowerSensor2 \cite{8366941}.

PMT library's core consists of a background thread to the profiled application that communicates and gathers power consumption information from the selected back end, such as NVML. 
PMT sampling frequency is dependent on the hardware and back end. For instance, NVML is able to sustain up to \SI{10}{ms} and RAPL up to \SI{500}{ms}.

PMT can be installed using CMake or a Spack \cite{gamblin_todd_2022_6973617,7830452} recipe \cite{schaap-spack}.
The library works in two modes: 1) \texttt{dump-mode} and 2) \texttt{measurement-mode}. The first one writes into a file timestamps and power measurements to be able to examine
the power consumption over time of an application. The second one simply provides the profiled code's average power consumption for quick energy-efficiency estimations.

\lstset{style=mystyle}
\begin{figure*}[!t]
\centering
\begin{minipage}{.56\textwidth}
\begin{lstlisting}[language=C, frame=tlrb, basicstyle=\scriptsize, caption=PMT example usage in C++ with NVML., label=codec]
#include <iostream>
#include <pmt/NVML.h>
#include <unistd.h>

int main(){
    std::unique_ptr<pmt::pmt> sensor(pmt::nvml::NVML::create());
    pmt::State start, end;
    start = sensor->read();
    sleep(5);
    end = sensor->read();
    std::cout << sensor->joules(start, end)  << " [J]" << std::endl;
    std::cout << sensor->watts(start, end)   << " [W]" << std::endl;
    std::cout << sensor->seconds(start, end) << " [S]" << std::endl;
}
\end{lstlisting}
\end{minipage}\hspace{0.5cm}
\begin{minipage}{.40\textwidth}
\vspace{0.56cm}
\begin{lstlisting}[language=Python, frame=tlrb,  basicstyle=\scriptsize, caption=PMT decorator example usage with NVML and rapl., label=codep]
import pmt
import time

@pmt.measure("rapl")
@pmt.measure("nvml")
def my_application():
    time.sleep(5)
    
if __name__ == "__main__":
    measures = my_application()
    for m in measures:
        print(m)
\end{lstlisting}
\end{minipage}
\end{figure*}

\begin{figure*}[!t]
    \centering
    \includegraphics[width=18cm]{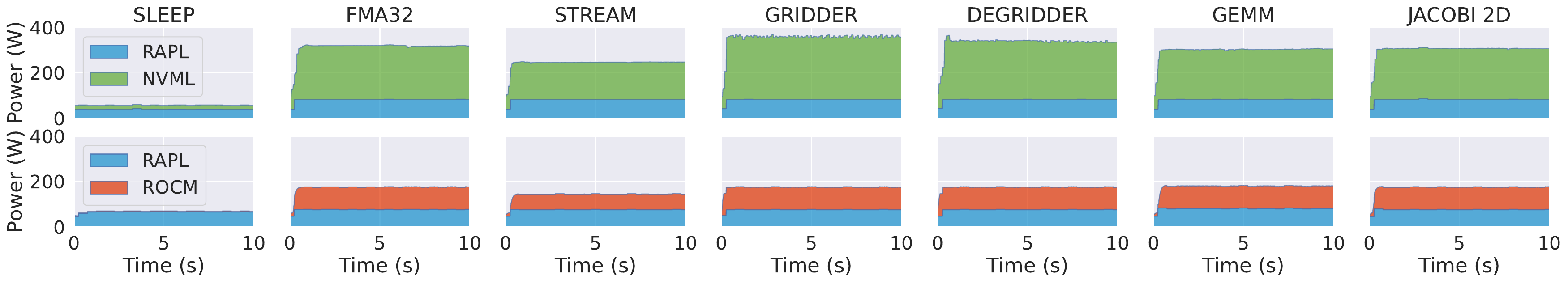}
    \caption{Benchmarking simple GPU kernels using PMT: GPUs power consumption (NVIDIA in green and AMD in red) is stacked on top of CPU one (in~blue).}
    \label{fig:powermeasurement}
\end{figure*}

PMT can be integrated into C++ and Python applications. In C++ the code must be instrumented as shown in \emph{Listing~\ref{codec}}.
Using PMT in a C++ application is a three-step process: 1) include the PMT header; 2) initialize PMT; 3) surround the region of interest with PMT activation and deactivation function calls.
This can be done similarly in Python by using Python bindings. However, we provide a simpler interface using Python decorators (see \emph{Listing~\ref{codep}}).
As shown in \emph{Listing~\ref{codep}}, the Python code only requires an \texttt{import} statement of the PMT library. In the case mentioned, we are collecting the measurement by using the NVML back end for NVIDIA GPUs and RAPL for AMD CPUs. The library support multiple decorators at the same time (see \emph{Listing} \ref{codep}). 

Typically, PMT has a small overhead in the order of \SI{1}{ms} in C++ and \SI{10}{ms} in Python. This overhead is cumulative when multiple decorators are used.

The \emph{Listings \ref{codec}} and \emph{\ref{codep}} shows the above-mentioned \texttt{measurement-mode} and how to programmatically query the measurement after the region of interest. Some changes are needed to activate the \texttt{dump-mode}. In Python, we just need to rename the decorator with \texttt{dump}. Similarly, in C++, we do not have to instantiate the start and end states. To start and stop this mode, we just need to surround some code with the respective APIs, e.g., \texttt{start\_dump\_thread}. In both Python and C++, the user must provide the filename where we would like to store the power measurements.

\section{Benchmarking}
\label{sec:benchmarking}

We monitor the power consumption on a system with an AMD Ryzen Threadripper 3970X CPU and two different GPUs: NVIDIA TITAN RTX and AMD Radeon PRO W6600. 
We run a set of simple GPU kernels to show in \emph{Fig. \ref{fig:powermeasurement}} PMT profiling capabilities: 1) a \texttt{SLEEP} kernel that leaves the GPU in idle; 2) \texttt{FMA32}, that executes many single-precision floating-point fused multiply and accumulate instructions; 3) \texttt{STREAM}, that stresses the GPU's device memory with a stream add motif; 4) \texttt{GRIDDER} and 5) \texttt{DEGRIDDER} are radio-astronimical gridder and degridder kernels inspired by \cite{9709826}; 6) \texttt{GEMM}, 7) \texttt{JACOBI2D} extracted from Polybench~\cite{6339595}.
We clearly notice in \emph{Fig. \ref{fig:powermeasurement}} that GPUs have low idle power consumption and can consume extremely high power while executing applications. 

In real-world scenarios, PMT can be used to assess application energy efficiency. This can be done in several ways. Users can extract measurements with PMT and derive energy efficiency metrics such as energy-delay product (EDP), which is the product of the execution time and the energy consumed, and the FLOPs efficiency, which can be expressed in GFLOP/s/W. Note that the last metric requires the number of FLOPs computed; this can be done by hand or using specific tools such as PAPI \cite{doi:10.1177/1094342019846287} and LIKWID \cite{5599200}.
\section{Related Work}
\label{sec:relatedwork}

Power meters remain the most accurate method to monitor the power consumption of current computing systems ~\cite{8366941}. However, it is not always possible and/or desirable to use a physical tool to measure power consumption. For quick and reliable estimates, built-in sensors can be used~\cite{10.1145/3387902.3392613}.
Most of the prior art focuses on a limited set of CPU and GPU brands ~\cite{7820633,10.1145/2962131,en12112204,9460501}, while PMT has more comprehensive hardware support.
PAPI~\cite{doi:10.1177/1094342019846287} and LIKWID~\cite{5599200} are noteworthy to be mentioned; they support power consumption monitoring for a limited set of architectures. 

To the best of our knowledge, PMT is the first library that provides a common interface to measure power consumption on numerous devices.
\section{Conclusion}
\label{sec:conclusion}


PMT allows power consumption measurements and monitoring with a simple interface on various hardware.
PMT's users can range from HPC application developers, whom could employ it to monitor and evaluate the energy efficiency of their code while optimizing it, to basic Python developers that would like to measure the power consumption of their application with a simple tool.
We plan to add support for upcoming hardware, such as Intel GPUs.
The library will be available at the link: \url{https://git.astron.nl/RD/pmt}.

\section*{Acknowledgment}
We thank anonymous reviewers of SC22 for their feedback and comments.
This research was partially funded by the Swiss State Secretariat for Education, Research and Innovation (SERI). We want to thank J. W. Romein (ASTRON) for its PowerSensor2 work, which lays the foundation of PMT.

\clearpage
\newpage
\bibliographystyle{IEEEtran}
\bibliography{IEEEabrv,ref_ext}



\end{document}